\documentclass[preprint,authoryear,12pt]{elsarticle}

\usepackage{amssymb}

\journal{Marine Pollution Bulletin}
\usepackage[dvipdfm]{hyperref}
\usepackage[active]{srcltx} 
\begin{document}

\begin{frontmatter}



\title{Index for Assessing Water Trophic Status in Semi-Enclosed Cuban Bays. Case Study: Cienfuegos Bay}


\author[1]{Mabel Seisdedo}
\address[1]{Centre of Environmental Studies of Cienfuegos\\
Calle 17 esq. Ave 46 s/n. Reparto Reina, Cienfuegos 55100, Cuba.
}
\author[1,2]{Roberto Herrera}
\address[2]{Department of Physics, University of Alberta. Edmonton, AB. T6G 2E1. Canada
}
\author[3]{Gustavo Arencibia}
\address[3]{Cuban Fisheries Research Centre. 5ta Ave y calle 246. Santa Fe. C.P.17100.
Ciudad de La Habana, Cuba.
}

\begin{abstract}
This paper aims at contributing to the coastal environmental management by developing a new trophic status index of the water (TSIW). The index is tailored to semi-enclosed bays with estuarine characteristic like the Cienfuegos bay in Cuba. We also propose pressure indicators related to exporting and assimilation capacities as a tool to assess the vulnerability of the system to eutrophication. The TSIW is based on response indicators to eutrophication processes showing correspondence with the predefined pressure indicators and previous reports on water quality. Thus, the proposed trophic status index is a reliable scientific tool to measure the current stage of the water quality and to establish a baseline for further studies.
\end{abstract}


\end{frontmatter}


\section{Introduction}
\label{intro}
Semi-enclosed bays often have estuarine characteristics due to the little exchange with marine waters and freshwater inputs from the rivers. Eutrophication of estuarine systems is a widespread environmental problem \citep{Cook2010}. Consequences of eutrophication can include high chlorophyll a (Chl.a) levels, anoxia and hypoxia events, and blooms of toxic and noxious algae \citep{Kennish2002,Rabalais2002,Beman2005}.

According to the conceptual contemporary model of the coastal eutrophication phase II \citep{Cloern2001}, the ecological impact of nutrient loading on a coastal system depends on some of the system attributes (e.g. morphology and residence time of the water body). This could be related to the inefficiency of applying trophic status indices for estuarine systems \citep{Coelho2007218,Salas2008} without taking into account differences among systems.
In Cuba, several studies have suggested eutrophication-like symptoms in semi-enclosed bays, such as high Chl.a. levels and toxic algae blooms events \citep{Reyes2008,Moreira2007}.

\cite{Areces1986} first reported the level of water eutrophication for the Cienfuegos bay using a global sense trophic status. More recently, \cite{Seisdedo2009} applying another method, classified the system with non-eutrophic water conditions. On the other hand, \cite{Reyes2008} with different approaches for assessing the trophic status of water in five Cuban bays found inconsistencies in the assessments.

Estuarine management activity relies on the assessments of water quality from values of trophic status. The trophic status of one system is compared to similar systems, as well as to its own evolution (e.g. changing pressures due to management actions). However, there is no reliable tool to assess the water trophic status in Cuban semi-enclosed bays. We propose new trophic status index based on response indicators of eutrophication.

The paper is organized as follow. First, we describe the location of our case study with the developing of the trophic status index. Then, we discuss our results providing the means to use the proposed index in Cuban semi-enclosed bays. Finally we discuss our results.

\section{Materials and Methods}
\label{Materials}
\subsection{Study area}
\begin{figure}[htb]
  \begin{center}
  \includegraphics[scale=0.6]{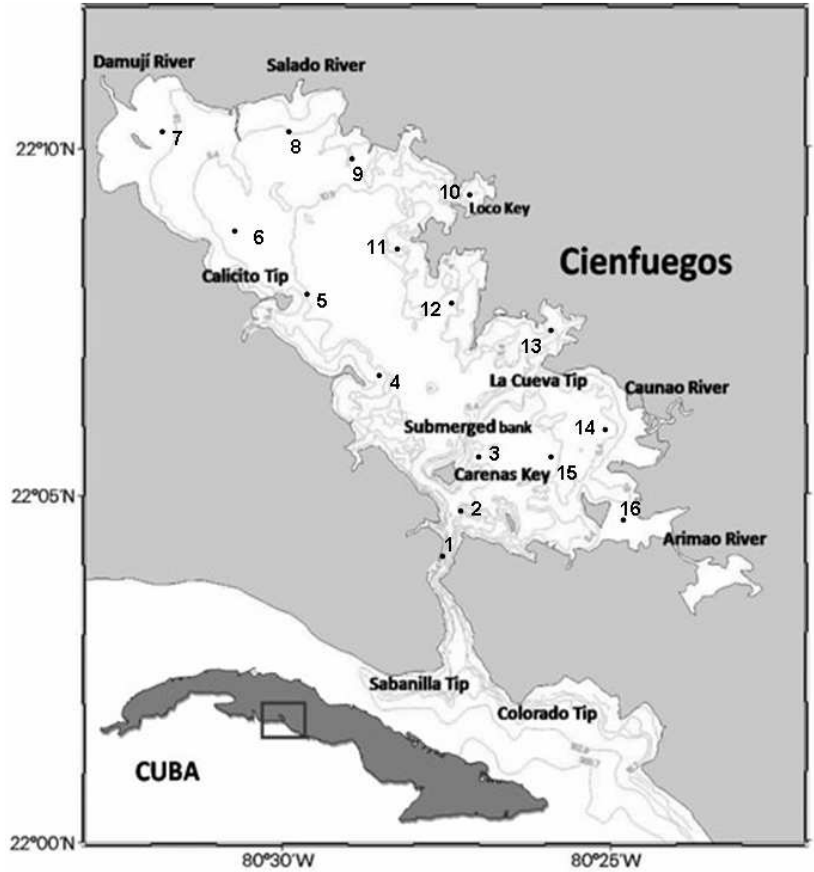}
 \end{center}
  \caption{Study area and sampling points location. Cienfuegos Bay, Cuba}\label{fig1}
\end{figure}

The Cienfuegos bay is situated in the southern central part of Cuba ($22^\circ 1'$~N, $80^\circ 20'$~W, see Figure \ref{fig1}). It is a semi-enclosed bay connected to the Caribbean Sea by a narrow channel 3.6~km long. Its area is 88.46~ km$^2$ and a total volume of 0.84~km$^3$ with an average depth of 9.5~m \citep{Alain2012}. The bay is divided into two natural lobes. The northern lobe suffers more anthropogenic impact: e.g. sewage discharges from the city and industrial area and the Salado and Damuj\'i rivers draining to the bay. The southern lobe is subjected to a lesser degree of pollution arriving from the Caonao and Arimao rivers.

\subsection{Developing trophic status index of water (\textrm{TSIW})}

The selection of the variables to include in the trophic status index (TSIW) is based on the frequency of appearance on previous studies (e.g.  \citep{Karydis2009,Vollenweider1998,Primpas2010}) and their applications (e.g. \citep{Coelho2007218,Reyes2008,FloresMontes2011}).  We consider 50 or more as a valuable frequency. Only response variables are selected since causal variables (nutrients) require specific criteria for each system  \citep{USEPA-822-B-01-0032001}.
The variable criteria for TSIW were based on the exposed levels indicating eutrophication in estuarine systems \citep{Bricker200339,USEPA-822-B-01-0032001,Cook2010}.
In order to combine the variables indicating status we follow \cite{CCME2001} and the number of indicators was defined considering the results of non parametric correlation analysis (Spearman) using selected variables from data of hydrographical surveys in Cienfuegos bay in 2009.

\subsection{Testing the \textrm{TSIW}}

Testing is a key stage in the development of any index to ensure that it is sensitive to environmental degradation and reliable, i.e. robust to the effects of natural variability \citep{Harrison2004}.The index sensitivity should be determined by testing its ability to correctly classify the health status of sites from an independent data set \citep{Lacouture2006,Astin2007895}. This may be by determining the correlation between index scores and independently-derived indices of environmental quality related to anthropogenic pressures \citep{Griffith2005,Romero2007,Uriarte2009}. On the other hand, the index reliability may be tested considering their differences among sites, seasons or years \citep{Brooks2009,Pyron2008,Martinho2008}.
The absence of annual nutrient load data limits the correlation analysis. This is the reason why in this study we define pressure indicators by analyzing their correspondence with the TSIW results. We apply our TSIW using hydrographical surveys data from Cienfuegos bay. Sampling was conducted on fourteen stations on February and September 2011; months were selected to correspond to dry and wet seasons, respectively. The sampling methods and chemical analysis follow \cite{APHA} and  \cite{UNEP1998} specifications.

We used expert criteria to define the pressure indicators (PI) based on some proposals \citep{USEPA-822-B-01-0032001,Bricker200339,Cardoso2004}. For the classification criteria we take into account the data collected on two semi-enclosed bays with similar physical and environmental characteristics \citep{Alain2012,Perez2008,Seisdedo2010}, and previous assessments of water trophic status \citep{Reyes2008,Seisdedo2010}.
We follow the methodology proposed by \cite{Cardoso2004}, which fits our analysis including a wide range of variables, such as morphology, temperature and salinity variability, uses and quality objectives.

\section{Results and discussion}
\label{results}
\subsection{Formulation of the TSIW}

\begin{figure}[htb]
  \begin{center}
  \includegraphics[scale=0.4]{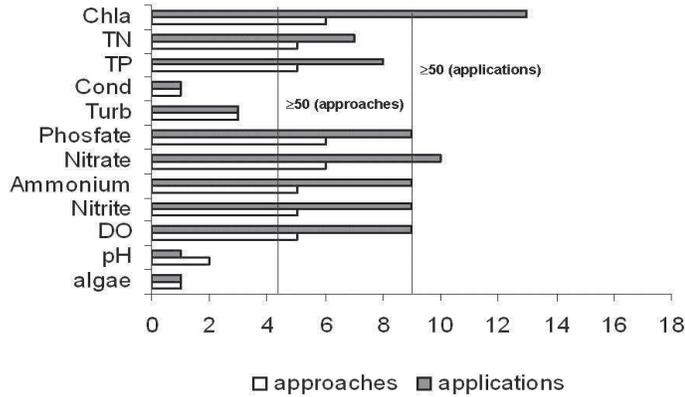}
 \end{center}
  \caption{Frequency of appearance of system variables in approaches of trophic status indices and their applications.}\label{fig2}
\end{figure}

Frequency analysis (Figure~\ref{fig2}), helps to select the independent variables Chl.a. and dissolved oxygen (DO), which are direct and indirect response variables, respectively (USEPA, 2001; Cloern, 2001). DO is considered hereafter as saturation percent (satDO, bottom), the same as in most of the applications. Figure~\ref{fig2} also shows that Chl.a stands out as the principal variable to use as a trophic status indicator \citep{Boyer2009} and the nutrients (causal variables) are used often in their different ways. On the other hand, we found no correlation between Chl.a. and satOD (r= -0.074, p=0.59, N=54). This result is in correspondence with \cite{USEPA-822-B-01-0032001}, and could be associated with temporal difference of the response of Chl.a. and satDO during the eutrophication process \citep{Cloern2001}.\\
\vspace{0.5cm}

Therefore we considered that eutrophication process could be described by two independent but complementary variables (Chl.a and satDO). Thus, the trophic status index of water is formulated as:

\begin{equation}
      TSIW = \frac{1}{\sqrt{2}} \sqrt{\textrm{DRI}^2 + \textrm{IRI}^2},
\end{equation}

\noindent where the direct response indicator is $\textrm{DRI} = c(\textrm{Chl.a})$, and the indirect response indicator is $\textrm{IRI} = c(\textrm{DO})$; with $c(x)$ the concentration of the variable $x$. $1/\sqrt{2}$ is a normalizing factor to bound the TSIW value between 0 and 1. TSIW reaches its maximum value when both DRI and IRI are 1.

DRI $= 0$ when $c(\textrm{Chl.a}) \leq 20~\mu g L^{-1}$; DRI = 1 when $c (\textrm{Chl.a}) > 20~\mu g L^{-1}$; IRI = 0 when $c(\textrm{satDO}) > 30~\%$; IRI = 1 when $c(\textrm{satDO}) \leq 30~\%$. Therefore, we proposed that quality criteria based on trophic status should consider satDO $ \leq 30~\%$, and Chl.a $ \geq 20~\mu g L^{-1}$.

\subsection{Selection of PI}

Table~\ref{tab:1} shows the classifications for both similar bays using criteria of two pressure indicators \citep{USEPA-822-B-01-0032001,Cardoso2004}. Only one of the indicators shows variability. This suggested the need for a better pressure indicator; in Table \ref{tab:2} we proposed a new indicator of assimilation capacity which considers two factors: dilution capacity and nutrient loads. This indicator describes better the variations on the nutrients and compares best with USEPA (2001).

\begin{table}[htbp]
  \centering
  \caption{Analysis of susceptibility factors in two similar Cuban semi-enclosed bays.}
\begin{tabular}{llll}
\hline
\multicolumn{1}{c}{Semi-enclosed} & \multicolumn{1}{c}{Dilution criteria } & \multicolumn{1}{c}{Criteria: exporting } & \multicolumn{1}{c}{Classification} \\
\multicolumn{1}{c}{ bays} & \multicolumn{1}{c}{ 1/$V_est$ (m$^3$)} & \multicolumn{1}{c}{capacity,  $t_{res}$ (days)} & \multicolumn{1}{c}{trophic status} \\
\hline
\hline
\multicolumn{1}{c}{Cienfuegos bay } & \multicolumn{1}{c}{Low ($10^{-9}$)} & \multicolumn{1}{c}{Mod-Low (39-50)} & \multicolumn{1}{c}{Non-eutrophic$^*$} \\
\hline
\multicolumn{1}{c}{Havana bay} & \multicolumn{1}{c}{Low ($10^{-8}$)} & \multicolumn{1}{c}{High (7 - 9)} & \multicolumn{1}{c}{Eutrophic} \\
\hline
\end{tabular}
$^*$ \small {Punctual eutrophic conditions in the season with low exporting capacity.}
  \label{tab:1}%
\end{table}%

\begin{table}[htbp]
  \centering
  \caption{Proposal of classification of new pressure indicator based on results of two similar Cuban semi-enclosed bays.}
   \begin{tabular}{lll}
    \hline
    \multicolumn{1}{c}{Semi-enclosed} & \multicolumn{1}{c}{Criteria: assimilation capacity} & \multicolumn{1}{c}{Classification} \\
    \multicolumn{1}{c}{ bays} & \multicolumn{1}{c}{ Q/$\textrm{V}_{est}$ (x $10^{-9}~\textrm{ton}/\textrm{m}^3$)} &  \\
    \hline
    \hline
    \multicolumn{1}{c}{Cienfuegos bay } & \multicolumn{1}{c}{$<$ 20 (2)$^*$} & \multicolumn{1}{c}{High} \\
    \hline
     \multicolumn{1}{c}{Havana bay} & \multicolumn{1}{c}{$\geq$ 20 (200)$^*$} & \multicolumn{1}{c}{Low} \\
     \hline
\end{tabular}
\\ $^*$(~): Assimilation capacity value for each Bay. 
  \label{tab:2}%
\end{table}%

\subsection{Application of TSIW}
During the dry season, IRI values were 0 in all sampling stations and IDI values were 0 except in the station 9, which showed the maxima value of Chl.a (see Table 3). During the wet season, only IRI values were 1 in two of the northern lobe stations (5 and 9). The northern lobe is known to have higher anthropogenic impact \citep{Tolosa2009} and there are sectors where the water exchange process is very slow \citep{Alain2012}. Previous studies have also reported signals of detrimental water quality in the northern lobe \citep{Areces1986,Seisdedo2006}. Although TSIW values of 0 prevailed (non-eutrophic condition) in both seasons, some stations showed eutrophic conditions.

\begin{table}[htbp]
  \centering
  \caption{Concentrations of selected variables, in Cienfuegos bay (2009-2010).}
\begin{tabular}{lll}
\hline
\multicolumn{1}{c}{Season} & \multicolumn{1}{c}{Interval of satDO (\%)} & \multicolumn{1}{c}{Interval of Chl.a (\%)} \\
\hline
\multicolumn{1}{c}{dry} & \multicolumn{1}{c}{57.4 - 103.1} & \multicolumn{1}{c}{0.72 - 38.2} \\
\hline
\multicolumn{1}{c}{wet} & \multicolumn{1}{c}{17.7 - 114.1} & \multicolumn{1}{c}{1.01 - 10.8} \\
\hline
\end{tabular}
  \label{tab:3}%
\end{table}%

Some authors relate eutrophication events to longer residence time of water (Herrera 2006; Jayachandran \& Bijoy 2012), however, our results showed that moderate exporting capacity also represents a big risk of eutrophication. This could be associated with the higher (i.e. doubled) nutrient concentrations reported by \cite{Seisdedo2009} in the wet season.
Finally, the trophic assessments using the TSIW showed correspondence with pressure indicators. Eutrophic conditions can be obtained in some areas when at least one of the pressure indicators is not high. The effectiveness of the index was confirmed by its correspondence with previous reports on water quality.

\section{Conclusions}
Semi-enclosed bays need specific trophic status indices to help improve the coastal environmental management. In this study we have demonstrated that TSIW is a reliable indicator of the water quality for the semi-enclosed Bay of Cienfuegos.
The indicators related to exporting and assimilation capacity are useful tools for testing the effectiveness of the trophic status index, as well as to assess the vulnerability of the system to eutrophication events.

\bibliographystyle{elsarticle-harv}
\bibliography{reftrophic}







\end{document}